\documentclass[prb,aps,preprint,showpacs,preprintnumbers,amsmath,amssymb,secnumroman,eqsecnum]{revtex4}

\usepackage{graphicx}
\usepackage{dcolumn}
\usepackage{bm}

\newcommand{\beq}{\begin{equation}}
\newcommand{\eeq}{\end{equation}}
\newcommand{\beqa}{\begin{eqnarray}}
\newcommand{\eeqa}{\end{eqnarray}}
\newcommand{\vc}[1]{\mbox{\boldmath $#1$}}

\newcommand{\vol}[1]{{\bf #1}}

\begin{document}


\title{Velocity relaxation of a porous sphere immersed in a viscous incompressible fluid}

\author{B. U. Felderhof}

 \email{ufelder@physik.rwth-aachen.de}
\affiliation{Institut f\"ur Theoretische Physik A\\ RWTH Aachen\\
Templergraben 55\\52056 Aachen\\ Germany\\
}%

\date{\today}

\begin{abstract}
Velocity relaxation of a spherically symmetric polymer, immersed in a viscous incompressible fluid, and after a sudden small impulse or a sudden twist from a state of rest, is studied on the basis of the linearized Navier-Stokes equations with an added Darcy type drag term. Explicit expressions for the translational and rotational velocity relaxation functions of the polymer and for the flow pattern of the fluid are derived for a uniform permeable sphere. Surprisingly, it is found that the added mass vanishes. For fairly large values of the ratio of sphere radius to the screening length characterizing the permeability the velocity relaxation functions in the short and intermediate time regime differ significantly from that of a sphere with no-slip boundary condition. At long times both relaxation functions show universal power law behavior.
\end{abstract}

\maketitle
\section{\label{I}Introduction}
The velocity of a solid sphere immersed in a viscous incompressible fluid after a sudden small impulse, starting from a state of rest, shows two conspicuous features as a function of time. At very short times the velocity is less than one would expect from its bare mass, due to added mass equal to half the mass of displaced fluid\cite{1}$^,$\cite{2}. At long times the velocity relaxation function decays with a $t^{-3/2}$ power law with a universal coefficient which is independent of the mass of the sphere, its radius, and the boundary condition at the surface\cite{3}$^,$\cite{4}. In contrast, the behavior at intermediate times does depend on these properties. For an understanding of the dynamics of polymer solutions it is of interest to study similarly the velocity relaxation function of a single polymer immersed in a viscous solvent. Throughout this paper we assume that the initial impulse is sufficiently small, so that low Reynolds number hydrodynamics applies.

Debye and Bueche\cite{5} first studied the steady-state friction coefficient of a polymer, modeled as a uniform permeable sphere immersed in a viscous fluid. The fluid flow was described by the Stokes equations with an added Darcy type term corresponding to the local friction caused by relative motion of polymer beads and fluid\cite{5}$^,$\cite{6}. It was found that in the limit of an impermeable sphere the friction coefficient tends to the Stokes value for a sphere with no-slip boundary condition. In earlier work we have generalized the calculation to a polymer model with more general radial dependence of the permeability\cite{7}. In the following we calculate the time-dependent velocity relaxation function of a uniform permeable sphere on the basis of the linearized Navier-Stokes equations with an added Darcy type drag term.

For values of the ratio of permeability screening length to sphere radius less than unity the velocity relaxation function differs significantly from that of a solid sphere with no-slip boundary condition. Surprisingly, in contrast to the latter case, the added mass is found to vanish, so that the polymer after the sudden impulse starts off with a much higher velocity than a sphere with no-slip boundary condition and the same radius and same bare mass.

If the screening length is small compared to the radius, the velocity decays in the initial stage rapidly to a much smaller value. The initial decay corresponds to a sharp peak in the relaxation spectrum. On the other hand, for large screening length the velocity relaxation is well described by a simple approximation corresponding to two poles of the admittance in the complex plane of the square root of frequency, determined by mass and steady-state translational friction coefficient. Then the relaxation spectrum is characterized by a single narrow peak and the relaxation function is nearly exponential.

We also study the relaxation of rotational velocity after a sudden small twist. Again the velocity relaxation function for a permeable sphere differs markedly from that of a solid sphere with no-slip boundary condition. If the screening length is small compared to the radius, the rotational velocity decays in the initial stage rapidly to a much smaller value, and again the initial decay corresponds to a sharp peak in the relaxation spectrum. On the other hand, for large screening length the velocity relaxation is well described by a simple approximation corresponding to two poles of the admittance in the complex plane of the square root of frequency, determined by moment of inertia and the steady-state rotational friction coefficient. Then the relaxation spectrum is characterized by a single narrow peak and the relaxation function is nearly exponential.

Other authors have attempted to evaluate the frequency-dependent friction coefficient of a porous sphere. Looker and Carrie\cite{7A} used a perturbative expansion for a slightly permeable sphere. Vainshtein and Shapiro\cite{7B} considered all values of the permeability, but their expression leads to an inconsistency at high permeability\cite{7C}. Ollila et al.\cite{7C} claim to perform an exact calculation for both a uniform sphere and a spherical shell, but their calculation is incorrect, as shown below. Ollila et al.\cite{7C} compare with lattice-Boltzmann calculations at a single frequency.

\section{\label{II}Flow equations}

We consider a spherically symmetric polymer, described in continuum approximation as a permeable sphere of radius $a$, immersed in a viscous incompressible fluid of shear viscosity $\eta$ and mass density $\rho$. For translational motion the flow velocity of the fluid $\vc{v}(\vc{r},t)$ and the pressure $p(\vc{r},t)$ are assumed to satisfy the linearized Navier-Stokes equations\cite{8}
\begin{equation}
\label{2.1}\rho\frac{\partial\vc{v}}{\partial
t}=\eta\nabla^2\vc{v}-\nabla
p-\eta\kappa^2(r)[\vc{v}-\vc{U}],\qquad\nabla\cdot\vc{v}=0,
\end{equation}
where the inverse length $\kappa(r)$ characterizes the local radially symmetric permeability, and $\vc{U}(t)$ is the time-dependent velocity of the polymer. In the following we assume $\kappa(r)$ to be uniform for $0<r<a$ and to vanish for $r>a$. The flow velocity $\vc{v}$ is assumed to vanish at infinity and the pressure $p$ tends to the constant ambient pressure $p_0$, so that the flow is driven by the sphere velocity $\vc{U}(t)$. We assume that the system is at rest for $t<0$ and is set in motion by an impulsive force acting at $t=0$. In order to calculate the friction exerted by the fluid on the sphere it suffices for the moment to assume that the velocity $\vc{U}(t)$ is known.

The solution of Eq. (2.1) is found most easily by a Fourier transform with respect to time. Thus we define Fourier components
$\vc{v}_\omega(\vc{r})$ and $p_\omega(\vc{r})$ by
\begin{equation}
\label{2.2}\vc{v}_\omega(\vc{r})=\int^\infty_0e^{i\omega
t}\vc{v}(\vc{r},t)\;dt,\qquad p_\omega(\vc{r})=\int^\infty_0
e^{i\omega t}p(\vc{r},t)\;dt.
\end{equation}
The equations for the Fourier components read
\begin{equation}
\label{2.3}\eta[\nabla^2\vc{v}_\omega-\alpha^2\vc{v}_\omega]-\nabla p_\omega=0,\qquad\nabla\cdot\vc{v}_\omega=0,\qquad\mathrm{for}\;r>a,
\end{equation}
with complex wavenumber
\begin{equation}
\label{2.4}\alpha=(-i\omega\rho/\eta)^{1/2},\qquad
\mbox{Re}\;\alpha>0.
\end{equation}
Inside the polymer
\begin{equation}
\label{2.5}\eta[\nabla^2\vc{v}_\omega-\beta^2\vc{v}_\omega]-\nabla p_\omega+\eta\kappa^2\vc{U}_\omega=0,\qquad\nabla\cdot\vc{v}_\omega=0,\qquad\mathrm{for}\;r<a,
\end{equation}
where
\begin{equation}
\label{2.6}\beta=\sqrt{\alpha^2+\kappa^2}.
\end{equation}
The equations must be solved under the conditions that $\vc{v}_\omega$ and $p_\omega$ tend to zero at infinity, and that the velocity and the normal-normal and normal-tangential components of the stress tensor are continuous at $r=a$. We choose coordinates such that the $z$ axis is in the direction of the impulsive applied force. The resulting polymer velocity is then also in the $z$ direction and the equations can be reduced to scalar form by the Ansatz
\begin{eqnarray}
\label{2.7}\vc{v}_\omega(\vc{r})&=&f_A(r)\vc{e}_z+f_B(r)(\vc{1}-3\hat{\vc{r}}\hat{\vc{r}})\cdot\vc{e}_z,\nonumber\\
p_\omega(\vc{r})&=&\eta g(r)\hat{\vc{r}}\cdot\vc{e}_z,
\end{eqnarray}
where the subscripts $A$ and $B$ refer to the two types of vector spherical harmonics that come into play\cite{9}. From Eqs. (A2) and (A5) of the Appendix we find that the radial functions must satisfy the coupled equations
\begin{eqnarray}
\label{2.8}\frac{d^2f_A}{dr^2}+\frac{2}{r}\frac{df_A}{dr}-\beta^2f_A-\frac{1}{3}\bigg(\frac{dg}{dr}+\frac{2}{r}\;g\bigg)=-\kappa^2U_\omega,\nonumber\\
\frac{d^2f_B}{dr^2}+\frac{2}{r}\frac{df_B}{dr}-\frac{6}{r^2}f_B-\beta^2f_B+\frac{1}{3}\bigg(\frac{dg}{dr}-\frac{1}{r}\;g\bigg)=0,\nonumber\\
\frac{d^2g}{dr^2}+\frac{2}{r}\frac{dg}{dr}-\frac{2}{r^2}\;g=0,\qquad\mathrm{for}\;r<a,
\end{eqnarray}
inside the polymer, and that outside they must satisfy
\begin{eqnarray}
\label{2.9}\frac{d^2f_A}{dr^2}+\frac{2}{r}\frac{df_A}{dr}-\alpha^2f_A-\frac{1}{3}\bigg(\frac{dg}{dr}+\frac{2}{r}\;g\bigg)=0,\nonumber\\
\frac{d^2f_B}{dr^2}+\frac{2}{r}\frac{df_B}{dr}-\frac{6}{r^2}f_B-\alpha^2f_B+\frac{1}{3}\bigg(\frac{dg}{dr}-\frac{1}{r}\;g\bigg)=0,\nonumber\\
\frac{d^2g}{dr^2}+\frac{2}{r}\frac{dg}{dr}-\frac{2}{r^2}\;g=0,\qquad\mathrm{for}\;r>a.
\end{eqnarray}
In spherical coordinates $(r,\theta,\varphi)$ the radial and polar components of the flow velocity are
\begin{equation}
\label{2.10}v_r(r,\theta)=(f_A-2f_B)\cos\theta,\qquad v_\theta(r,\theta)=-(f_A+f_B)\sin\theta,
\end{equation}
and $v_\varphi=0$. The relevant components of the stress tensor $\vc{\sigma}_\omega=\eta(\nabla\vc{v}_\omega+\widetilde{\nabla\vc{v}_\omega})-p_\omega\vc{1}$ are
\begin{eqnarray}
\label{2.11}\sigma_{rr}(r,\theta)&=&\eta\big(2f'_A-4f'_B-g\big)\cos\theta,\nonumber\\
\sigma_{r\theta}(r,\theta)&=&\sigma_{\theta r}(r,\theta)=\eta\big(-f'_A-f'_B+\frac{3}{r}f_B\big)\sin\theta,
\end{eqnarray}
where the prime denotes the derivative with respect to $r$.

The solution of Eq. (2.8) takes the form
\begin{eqnarray}
\label{2.12}f_A(r)&=&2A_Ni_0(\beta r)+A_P,\qquad f_B(r)=A_Ni_2(\beta r),\nonumber\\
g(r)&=&-A_P\beta^2 r+\kappa^2 U_\omega r,\qquad\mathrm{for}\;r<a,
\end{eqnarray}
with coefficients $A_N,A_P$ and modified spherical Bessel functions\cite{10} $i_l(z)$.
The solution of Eq. (2.9) takes the form\cite{11}
\begin{eqnarray}
\label{2.13}f_A(r)&=&2B_Nk_0(\alpha r),\qquad f_B(r)=B_Nk_2(\alpha r)-B_P/r^3,\nonumber\\
g(r)&=&B_P\frac{\alpha^2}{r^2},\qquad\mathrm{for}\;r>a,
\end{eqnarray}
with coefficients $B_N,B_P$ and modified spherical Bessel functions $k_l(z)$.

The requirement that $v_r,v_\theta,\sigma_{rr},\sigma_{r\theta}$ are continuous at $r=a$ leads to four equations for the four coefficients. These have the solution
\begin{eqnarray}
\label{2.14}A_N&=&-\frac{\alpha k_1}{D}\;\kappa^2U_\omega,\qquad A_P=2\frac{\alpha i_0k_1+\beta i_1k_0}{D}\;\kappa^2U_\omega,\nonumber\\
B_N&=&\frac{\beta i_1}{D}\;\kappa^2U_\omega,\qquad B_P=a^3\frac{\alpha i_2k_1+\beta i_1k_2}{D}\;\kappa^2U_\omega,
\end{eqnarray}
with denominator
 \begin{equation}
\label{2.15}D=\alpha^3i_2k_1+\alpha^2\beta i_1k_2+2\alpha\beta^2i_0k_1+2\beta^3i_1k_0,
\end{equation}
and abbreviations
 \begin{equation}
\label{2.16}i_l=i_l(\beta a),\qquad k_l=k_l(\alpha a).
\end{equation}
It is seen that as a consequence $f_A(r),f_B(r),g(r)$ are all continuous at $r=a$. It may be shown that in the limit $\omega\rightarrow 0$ the solution tends to the known steady-state flow pattern\cite{7}.

\section{\label{III}Polymer velocity}

From the flow pattern calculated above we can evaluate the hydrodynamic force exerted by the fluid on the sphere. This is given by the integral of the friction between the polymer and the fluid
 \begin{equation}
\label{3.1}\vc{K}_{p\omega}=\int_{r<a}\eta\kappa^2(\vc{v}_\omega-\vc{U}_\omega)\;d\vc{r}.
\end{equation}
From the linearized Navier-Stokes equations Eq. (2.1) it follows that this can be expressed alternatively as
 \begin{equation}
\label{3.2}\vc{K}_{p\omega}=\int_{S(a+)}\vc{\sigma}_\omega\cdot\hat{\vc{r}}\;dS+i\omega\rho\int_{r<a}\vc{v}_\omega\;d\vc{r},
\end{equation}
where the first term is the integral over the stress tensor over a spherical surface just outside the polymer.
From Eqs. (2.11) and (2.13) we find $\vc{K}_{p\omega}=K_{p\omega}\vc{e}_z$ with
 \begin{equation}
\label{3.3}K_{p\omega}=-\frac{4\pi}{3}\eta\bigg[6\alpha a^2k_1B_N+\alpha^2B_P+\alpha^2a^2\big(6A_Ni_1/\beta+A_P a\big)\bigg].
\end{equation}
From Eq. (3.1) we have the alternative expression
  \begin{equation}
\label{3.4}K_{p\omega}=\frac{4\pi}{3}\eta\kappa^2a^2\big[6A_Ni_1/\beta+A_P a-aU_\omega\big].
\end{equation}
The translational friction coefficient $\zeta_T(\omega)$ is defined by
 \begin{equation}
\label{3.5}K_{p\omega}=-\zeta_T(\omega)U_\omega.
\end{equation}
As a check we find from the above expressions in the steady-state limit $\omega\rightarrow 0$
 \begin{equation}
\label{3.6}\zeta_T(0)=6\pi\eta aZ_0(\sigma),\qquad\sigma=\kappa a,
\end{equation}
where
\begin{equation}
\label{3.7}Z_0(\sigma)=\frac{2\sigma^2G_0(\sigma)}{2\sigma^2+3G_0(\sigma)},\qquad G_0(\sigma)=1-\frac{\tanh\sigma}{\sigma},
\end{equation}
in agreement with the result of Debye and Bueche\cite{5}.

For small $\alpha$ the friction coefficient behaves as
\begin{equation}
\label{3.8}\zeta_T(\omega)=\zeta_T(0)+\frac{1}{6\pi\eta}\;\zeta_T(0)^2\alpha+O(\alpha^2).
\end{equation}
At high frequency the friction coefficient tends to a constant given by
\begin{equation}
\label{3.9}\zeta_{T\infty}=\lim_{\omega\rightarrow\infty}\zeta_T(\omega)=\frac{4\pi}{3}\eta\kappa^2a^3.
\end{equation}
This corresponds to instantaneous friction against the polymer beads.

The calculation by Ollila et al.\cite{7C} of the frequency-dependent friction coefficient, which they claim to be exact, is wrong. Their Eq. (3.26) is incorrect, as is seen by a comparison of Eqs. (3.1) and (3.2). They leave out the last term in Eq. (3.2).

The sphere velocity is determined by the equation of motion
\begin{equation}
\label{3.10}-i\omega m_p\vc{U}_\omega=\vc{K}_{p\omega}+\vc{E}_\omega,
\end{equation}
where $m_p$ is the polymer mass and $\vc{E}_\omega$ is the external mechanical force applied to the polymer. Hence we find
\begin{equation}
\label{3.11}\vc{U}_\omega=\mathcal{Y}_T(\omega)\vc{E}_\omega,
\end{equation}
where $\mathcal{Y}_T(\omega)$ is the translational admittance, given by
\begin{equation}
\label{3.12}\mathcal{Y}_T(\omega)=[-i\omega m_p+\zeta_T(\omega)]^{-1}.
\end{equation}
In general the added mass of the particle is defined by
\begin{equation}
\label{3.13}m_a=\lim_{\omega\rightarrow\infty}\frac{\zeta_T(\omega)}{-i\omega}.
\end{equation}
In the present case $m_a=0$. We note that
\begin{equation}
\label{3.14}\lim_{\kappa\rightarrow\infty}\zeta_T(\omega)=6\pi\eta a\bigg(1+\alpha a+\frac{1}{3}\alpha^2a^2\bigg).
\end{equation}
The last term in brackets differs from that for a sphere with no-slip boundary condition, where it is replaced by $\frac{1}{9}\alpha^2a^2$, corresponding to $m_a=\frac{1}{2}m_f$, where $m_f=(4\pi/3)\rho a^3$ is the mass of fluid displaced by the sphere. This no-slip result follows from just the first term in Eq. (3.2), denoted earlier\cite{11}$^,$\cite{11B} as $\vc{K}_\omega$. In the present case the incoming momentum flow is used partly as a force acting on the polymer, partly it is used to accelerate the fluid within the sphere of radius $a$. The expression Eq. (3.14) shows that the limit of large $\kappa a$ is tricky. The last term suggests an added mass, even though at any finite value of $\kappa a$ the added mass vanishes.

We consider in particular the applied force
\begin{equation}
\label{3.15}\vc{E}(t)=\vc{P}\;\delta(t),
\end{equation}
where $\vc{P}$ is the imparted impulse. Correspondingly $\vc{E}_\omega=\vc{P}$. We define the translational velocity relaxation function $\gamma_T(t)$ by
\begin{equation}
\label{3.16}U(t)=\frac{P}{m_p}\gamma_T(t),\qquad t>0.
\end{equation}
It has the properties
\begin{equation}
\label{3.17}\gamma_T(0+)=1,\qquad\int^\infty_0\gamma_T(t)\;dt=\frac{m_p}{\zeta_T(0)}.
\end{equation}
Since the added mass vanishes, here the bare mass $m_p$ appears, rather than the effective mass $m^*=m_p+\frac{1}{2}m_f$, as for a sphere with no-slip or mixed slip-stick boundary condition\cite{3}$^,$\cite{11A}.
The relaxation function is related to the admittance by
\begin{equation}
\label{3.18}\int^\infty_0e^{i\omega t}\gamma_T(t)\;dt=m_p\mathcal{Y}_T(\omega).
\end{equation}
Hence the relaxation function $\gamma_T(t)$ can be evaluated by inverse Fourier transform.

Corresponding to Eq. (3.8) the admittance has the low frequency expansion
\begin{equation}
\label{3.19}\mathcal{Y}_T(\omega)=\mu_T(0)-\frac{\alpha}{6\pi\eta}+O(\omega),
\end{equation}
where $\mu_T(0)=1/\zeta_T(0)$ is the steady-state mobility. The second term gives rise to the long-time behavior of the velocity,
\begin{equation}
\label{3.20}\vc{U}(t)\approx\frac{1}{12\rho(\pi\nu t)^{3/2}}\;\vc{P}\qquad\mathrm{as}\;t\rightarrow\infty,
\end{equation}
where $\nu=\eta/\rho$ is the kinematic viscosity. It is remarkable that this is independent of the nature of the polymer, and depends only on the properties of the fluid\cite{12}.

\section{\label{IV}Analysis of velocity relaxation}

It is worthwhile to analyze the behavior of the velocity relaxation function in some more detail. The above expressions show that the friction coefficient is conveniently regarded as a function of the complex variable $x=\alpha a$. It is convenient to define the dimensionless admittance $\hat{F}_T(x)$ as
\begin{equation}
\label{4.1}\hat{F}_T(x)=\frac{4\pi m_p}{3m_f}\;\eta a\mathcal{Y}_T(\omega),
\end{equation}
where we have chosen the prefactor such that $\hat{F}_T(x)$ behaves as $1/x^2$ for large $x$. The function takes the form
\begin{equation}
\label{4.2}\hat{F}_T(x)=\frac{1}{x^2+MZ_T(x)},
\end{equation}
where we have abbreviated
\begin{equation}
\label{4.3}M=\frac{9m_f}{2m_p},\qquad\zeta_T(\omega)=6\pi\eta a Z_T(x).
\end{equation}

As in previous analysis\cite{13} we write the reduced admittance as a sum of simple poles in the complex $x$ plane of the form
\begin{equation}
\label{4.4}\hat{F}_T(x)=\sum_j\frac{A_j}{x-x_j}.
\end{equation}
Since the function behaves as $1/x^2$ at large $x$ one has the sum rules
\begin{equation}
\label{4.5}\sum_jA_j=0,\qquad\sum_jA_jx_j=1.
\end{equation}
In addition we find from Eq. (3.8)
\begin{equation}
\label{4.6}\sum_j\frac{A_j}{x_j}=\frac{-1}{MZ_0},\qquad\sum_j\frac{A_j}{x_j^2}=\frac{1}{M},
\end{equation}
where $Z_0=Z_T(0)$ is given by Eq. (3.7). Finally, Eq. (3.9) implies the sum rules
\begin{equation}
\label{4.7}\sum_jA_jx_j^2=0,\qquad\sum_jA_jx_j^3=-MZ_{0\infty},\qquad Z_{0\infty}=\frac{2}{9}\kappa^2 a^2,
\end{equation}
where $Z_{0\infty}$ corresponds to $\zeta_{T\infty}$. The relaxation function is given by
\begin{equation}
\label{4.8}\gamma_T(t)=\sum_jA_jx_jw(-ix_j\sqrt{t/\tau_v}),
\end{equation}
where $w(z)$ is the $w$ function\cite{10} $w(z)=\exp(-z^2)\mathrm{erfc}(-iz)$, and $\tau_v=a^2/\nu$ is the viscous relaxation time. The second sum rule in Eq. (4.5) corresponds to the initial value $\gamma_T(0+)=1$.

We write the function $\hat{F}_T(x)$ in the form\cite{14}
\begin{equation}
\label{4.9}\hat{F}_T(x)=\frac{1}{MZ_0+MZ_0^2x+x^2+x^2\psi(x)},
\end{equation}
with a function $\psi(x)$ which tends to zero for large $x$. If we approximate $\psi(x)$ as a ratio of two polynomials $A(x)$ and $B(x)$ of degree $d-3$ and $d-2$ respectively, then $\hat{F}_T(x)$ is approximated by a Pad\'e approximant $\hat{F}_{Pd}(x)$ with $d$ poles. The Pad\'e approximant is obtained by choosing the polynomials such that exact values are obtained at $2d-4$ selected points on the positive $x$ axis.
The quality of the approximation can be gauged by comparison of the corresponding $\gamma_{TPd}(t)$ with $\gamma_T(t)$ as calculated by numerical Fourier inversion of the exact $\hat{F}_T(x)$, and by comparison with the sum rules (4.7). The sum rules in Eqs. (4.5) and (4.6) are satisfied automatically by construction. The function $\gamma_{TPd}(t)$ provides a smooth interpolation between the exact short-time and long-time behavior.

For large values of $\kappa a$ the friction coefficient at zero frequency tends to the Stokes value $6\pi\eta a$ for a no-slip sphere. It is therefore of interest to compare the relaxation function for large $\kappa a$ with that for a no-slip sphere. The latter follows from
\begin{equation}
\label{4.10}\hat{F}_{Tns}(x)=\frac{1}{M+Mx+(m^*/m_p)x^2},
\end{equation}
with effective mass $m^*=m_p+\frac{1}{2}m_f$. This yields the relaxation function\cite{15}$^-$\cite{18}
 \begin{equation}
\label{4.11}\gamma_{Tns}(t)=\frac{m_p}{m^*}\;\frac{1}{y_+-y_-}\;\big[y_+w(-iy_+\sqrt{t/\tau_{Mn}})-y_-w(-iy_-\sqrt{t/\tau_{Mn}})\big],
\end{equation}
with
 \begin{equation}
\label{4.12}y_\pm=-\frac{1}{2}\sqrt{M^*}\pm\frac{1}{2}\sqrt{M^*-4},\qquad M^*=\frac{9m_f}{2m^*},\qquad\tau_{Mn}=\frac{m^*}{6\pi\eta a}.
\end{equation}
The function has initial value $\gamma_{Tns}(0+)=m_p/m^*$, and it has the same long-time tail as $\gamma_T(t)$, given by $m_p/[12\rho(\pi\nu t)^{3/2}]$, as follows from Eqs. (3.16) and (3.20).

The simplest Pad\'e approximant is obtained by neglecting the function $\psi(x)$ in Eq. (4.9). This yields
\begin{equation}
\label{4.13}\hat{F}_{TP2}(x)=\frac{1}{MZ_0+MZ_0^2x+x^2}.
\end{equation}
The corresponding relaxation function is
\begin{equation}
\label{4.14}\gamma_{TP2}(t)=A_+x_+w(-ix_+\sqrt{t/\tau_v})+A_-x_-w(-ix_-\sqrt{t/\tau_v}),
\end{equation}
with the values
\begin{equation}
\label{4.15}A_\pm=\frac{\pm 1}{x_+-x_-},\qquad x_\pm=-\frac{1}{2}MZ_0^2\pm\frac{1}{2}\sqrt{MZ_0(MZ_0^3-4)}.
\end{equation}
We call this the two-pole approximation.

In our numerical work we consider a neutrally buoyant polymer with $m_f=m_p$, so that $M=9/2$ and $M^*=3$. We shall consider three values of the parameter $\kappa a$, namely $\kappa a=0.2,\;\kappa a=1$, and $\kappa a=5$. For these three values we find for the roots $x_\pm$
\begin{eqnarray}
\label{4.16}x_\pm&=&-0.0002\pm0.199i,\qquad\kappa a=0.2,\nonumber\\
x_\pm&=&-0.069\pm0.886i,\qquad\kappa a=1,\nonumber\\
x_\pm&=&-1.311\pm1.311i,\qquad\kappa a=5,
\end{eqnarray}
showing considerable variation as a function of $\kappa a$. It turns out that for small $\kappa a$ the two-pole approximation is quite accurate, but for larger values a larger number of poles is needed for an accurate description. For a neutrally buoyant no-slip sphere we have $\tau_v/\tau_{Mn}=3$ and find
\begin{equation}
\label{4.17}y_\pm\sqrt{\tau_v/\tau_{Mn}}=-\frac{3}{2}\pm\frac{1}{2}i\sqrt{3}=-1.5\pm0.866i.
\end{equation}
The no-slip relaxation function starts at $\gamma_{Tns}(0+)=2/3$, whereas $\gamma_T(0+)=1$. In Fig. 1 we show $\log_{10}\gamma_T(t)$ as a function of $\log_{10}(t/\tau_v)$ for the three different values of $\kappa a$, and compare with the no-slip function $\gamma_{Tns}(t)$. In this plot the no-slip function can hardly be distinguished from the one for $\kappa a=5$. However, for short and intermediate times the two functions are actually quite different, as shown in Fig. 2. At $t=2\tau_v$ the functions have decayed to $\gamma_T(2\tau_v)=0.023$ and $\gamma_{Tns}(2\tau_v)=0.017$. In the long-time regime the two functions become identical.

The relaxation function may be expressed as
\begin{equation}
\label{4.18}\gamma_T(t)=\int^\infty_0p_T(u)e^{-ut/\tau_v}\;du,
\end{equation}
with a spectral density $p_T(u)$ which has been normalized to
\begin{equation}
\label{4.19}\int^\infty_0p_T(u)\;du=1.
\end{equation}
The spectral density can be found from the exact solution by use of the rule\cite{18A}
\begin{equation}
\label{4.20}p_T(u)=-\frac{1}{\pi}\;\mathrm{Im}\;\hat{F}_T(x=i\sqrt{u}).
\end{equation}
For the no-slip case the expression for the spectral density reads\cite{17}
\begin{equation}
\label{4.21}p_{Tns}(u)=\frac{m_p}{\pi m^*}\;\frac{{M^*}\sqrt{u}}{{M^*}^2+M^*(M^*-2)u+u^2},
\end{equation}
where we have used $M=m^*M^*/m_p$. In Fig. 3 we show the spectral density $p_T(u)$ for $\kappa a=5$ as a function  of $\log_{10}u$, and compare with the spectral density $p_{Tns}(u)$ for the neutrally buoyant no-slip sphere with $M^*=3$. We also compare with the spectral density corresponding to a six-pole Pad\'e expression obtained from a fit to the exact $\mathcal{Y}_T(\omega)$ at eight positive values of $\alpha a$. With the Pad\'e approximant we find for the first sum in Eq. (4.7) the value $0.003$ instead of zero, and for the second sum $-24.69$ instead of $-25$. It is seen that the six-pole expression reproduces the spectral density quite well. There is a sharp peak for large values of $u$ which is missing from the spectral density $p_{Tns}(u)$. The peak corresponds to two zeroes of the denominator of $\hat{F}_T(x)$ at $-0.010\pm6.549i$. The broad peak at smaller values of $u$ corresponds to zeroes at $-0.829\pm1.156i$. The spectral density $p_T(u)$ is positive, which shows that the relaxation function $\gamma_T(t)$ is completely monotone\cite{19}. The relaxation function is closely approximated by the function $\gamma_{TP6}(t)$ obtained from the Pad\'e approximant, and given by a sum of six terms of the form Eq. (4.8). In Fig. 4 we show the ratio of the two functions as a function of $\log_{10}(t/\tau_v)$. The exact function $\gamma_T(t)$ is obtained by numerical computation of the Fourier transform of the admittance.

For yet larger values of $\kappa a$ qualitatively the same picture obtains. As $\kappa a$ increases the sharp peak in the spectral density moves to the right, corresponding to faster relaxation.

For small values of $\kappa a$ the admittance is well approximated by the two-pole Pad\'e approximant given by Eq. (4.13). For $\kappa a=0.2$ the ratio of the exact relaxation function $\gamma_T(t)$ and the approximate $\gamma_{TP2}(t)$ is between $0.998$ and unity over the whole range of time. This also shows that the spectral density is well approximated by a single sharp peak. The spectral density in the two-pole approximation, corresponding to Eq. (4.13), is\cite{17}
\begin{equation}
\label{4.20}p_{TP2}(u)=\frac{1}{\pi}\;\frac{{MZ_0^2}\sqrt{u}}{M^2Z_0^2+MZ_0(MZ_0^3-2)u+u^2}.
\end{equation}
In Fig. 5 we compare the exact spectral density with the two-pole approximation for $\kappa a=0.2$. The peak is quite sharp, corresponding to the small value of the real part in the first line of Eq. (4.16). The corresponding pole of the exact admittance is at $-0.0002+0.2i$. The sharpness of the peak implies that the relaxation function is nearly exponential.

\section{\label{V}Rotational motion}

Next we consider rotational motion due to a sudden twist, corresponding to a time-dependent torque $\vc{N}(t)=\vc{L}\delta(t)$, applied to the permeable sphere of moment of inertia $I_p$, causing the sphere to rotate and the fluid to move. The torque will be assumed small, so that we can again use linearized equations of motion. We shall be interested in calculating the time-dependent rotational velocity $\vc{\Omega}(t)$ of the sphere, as well as the corresponding flow pattern of the fluid.

We define Fourier components of the rotational velocity by
\begin{equation}
\label{5.1}\vc{\Omega}_\omega=\int^\infty_0e^{i\omega t}\vc{\Omega}(t)\;dt.
\end{equation}
The pressure remains constant and uniform, so that the equations for the Fourier components of the flow velocity read
\begin{equation}
\label{5.2}\eta[\nabla^2\vc{v}_\omega-\alpha^2\vc{v}_\omega]=0,\qquad\nabla\cdot\vc{v}_\omega=0,\qquad\mathrm{for}\;r>a.
\end{equation}
Inside the polymer
\begin{equation}
\label{5.3}\eta[\nabla^2\vc{v}_\omega-\beta^2\vc{v}_\omega]+\eta\kappa^2\vc{\Omega}_\omega\times\vc{r}=0,\qquad\nabla\cdot\vc{v}_\omega=0,\qquad\mathrm{for}\;r<a.
\end{equation}
Since the pressure remains uniform the same equations apply for rotational motion in a compressible fluid\cite{13}.

The equations can be reduced to scalar form by the Ansatz
\begin{equation}
\label{5.4}\vc{v}_\omega(\vc{r})=f_C(r)\vc{e}_z\times\hat{\vc{r}}.
\end{equation}
By use of Eq. (A5) of the Appendix we find that the radial function $f_C(r)$ must satisfy the equation
\begin{eqnarray}
\label{5.5}\frac{d^2f_C}{dr^2}+\frac{2}{r}\frac{df_C}{dr}-\frac{2}{r}f_C-\beta^2f_C=-\kappa^2\Omega_\omega r,\qquad\mathrm{for}\;r<a,\nonumber\\
\frac{d^2f_C}{dr^2}+\frac{2}{r}\frac{df_C}{dr}-\frac{2}{r}f_C-\alpha^2f_C=0,\qquad\mathrm{for}\;r>a.
\end{eqnarray}
In spherical coordinates the only non-vanishing component of the flow velocity is
\begin{equation}
\label{5.6}v_\varphi(r)=f_C(r)\sin\theta.
\end{equation}
The relevant components of the stress tensor $\vc{\sigma}_\omega=\eta(\nabla\vc{v}_\omega+\widetilde{\nabla\vc{v}_\omega})$ are
\begin{equation}
\label{5.7}\sigma_{r\varphi}(r)=\sigma_{\varphi r}(r)=\eta\bigg(f'_C-\frac{f_C}{r}\bigg)\sin\theta.
\end{equation}
The solution of Eq. (5.5) takes the form
\begin{eqnarray}
\label{5.8}f_C(r)&=&A_Mi_1(\beta r)+\frac{\kappa^2}{\beta^2}\;\Omega_\omega r,\qquad\mathrm{for}\;r<a,\nonumber\\
&=&B_Mk_1(\alpha r),\qquad\mathrm{for}\;r>a.
\end{eqnarray}
From the conditions that $v_\varphi$ and $\sigma_{r\varphi}$ be continuous at $r=a$ we find for the coefficients
\begin{equation}
\label{5.9}A_M=-\frac{\kappa^2}{\beta^2}\frac{3k_1+\alpha ak_0}{\alpha i_1k_0+\beta i_0k_1}\;\Omega_\omega,\qquad B_M=-\frac{\kappa^2}{\beta^2}\frac{3i_1-\beta ai_0}{\alpha i_1k_0+\beta i_0k_1}\;\Omega_\omega.
\end{equation}
It may be shown that in the limit $\omega\rightarrow 0$ the solution tends to the known steady-state flow pattern\cite{26}.

\section{\label{VI}Polymer rotational velocity}

The hydrodynamic torque exerted by the fluid on the polymer is given by the integral
 \begin{equation}
\label{6.1}\vc{T}_{p\omega}=\int_{r<a}\eta\kappa^2\vc{r}\times(\vc{v}_\omega-\vc{\Omega_\omega\times\vc{r}})\;d\vc{r}.
\end{equation}
From the linearized equations Eq. (2.1) it follows that this can be expressed alternatively as
 \begin{equation}
\label{6.2}\vc{T}_{p\omega}=\int_{S(a+)}\vc{r}\times(\vc{\sigma}_\omega\cdot\hat{\vc{r}})\;dS+i\omega\rho\int_{r<a}\vc{r}\times\vc{v}_\omega\;d\vc{r}.
\end{equation}
This yields $\vc{T}_{p\omega}=T_{p\omega}\vc{e}_z$ with
 \begin{equation}
\label{6.3}T_{p\omega}=-\frac{8\pi}{3}\eta\bigg[a^2(\alpha ak_0+3k_1)B_M+\frac{\alpha^2a^3}{\beta}g_2A_M+\frac{\alpha^2\kappa^2}{5\beta^2}a^5\Omega_\omega\bigg].
\end{equation}
From Eq. (6.1) we have the alternative expression
 \begin{equation}
\label{6.4}T_{p\omega}=\frac{8\pi}{3}\eta\kappa^2a^3\bigg[\frac{g_2}{\beta}A_M-\frac{\alpha^2a^2}{5\beta^2}\Omega_\omega\bigg].
\end{equation}
The rotational friction coefficient $\zeta_R(\omega)$ is defined by
 \begin{equation}
\label{6.5}T_{p\omega}=-\zeta_R(\omega)\Omega_\omega.
\end{equation}
As a check we find from the above expressions in the steady-state limit $\omega\rightarrow 0$
 \begin{equation}
\label{6.6}\zeta_R(0)=8\pi\eta a^3Z_0(\sigma),\qquad Z_0(\sigma)=1+\frac{3}{\sigma^2}-\frac{3}{\sigma}\coth \sigma,\qquad\sigma=\kappa a,
\end{equation}
in agreement with the result of Felderhof and Deutch\cite{20}.

For small $\alpha$ the friction coefficient behaves as
\begin{equation}
\label{6.7}\zeta_R(\omega)=\zeta_R(0)+\zeta_{R2}\alpha^2a^2-\frac{1}{24\pi\eta}\;\zeta_R(0)^2\alpha^3+O(\alpha^4),
\end{equation}
where $\zeta_{R2}$ is given by the complicated expression
\begin{equation}
\label{6.8}\zeta_{R2}=\frac{2}{5}\pi\eta a^3\frac{1}{\sigma^4(\sinh\sigma)^2}\big[60+60\sigma^2-4\sigma^4-(60-30\sigma^2-4\sigma^4)\cosh2\sigma+5\sigma(3-4\sigma^2)\sinh2\sigma\big],
\end{equation}
with the behavior
\begin{equation}
\label{6.9}\zeta_{R2}=\frac{16}{945}\pi\eta a^3\sigma^4+O(\sigma^5),\qquad\lim_{\sigma\rightarrow\infty}\zeta_{R2}=\frac{16}{5}\pi\eta a^3.
\end{equation}
At high frequency the friction coefficient tends to a constant given by
\begin{equation}
\label{6.10}\zeta_{R\infty}=\frac{8\pi}{15}\eta\kappa^2a^5.
\end{equation}

The rotational velocity of the polymer is determined by the equation of motion
\begin{equation}
\label{6.11}-i\omega I_p\vc{\Omega}_\omega=\vc{T}_{p\omega}+\vc{N}_\omega,
\end{equation}
where $\vc{N}_\omega$ is the external mechanical torque applied to the polymer. Hence we find
\begin{equation}
\label{6.12}\vc{\Omega}_\omega=\mathcal{Y}_R(\omega)\vc{N}_\omega,
\end{equation}
where $\mathcal{Y}_R(\omega)$ is the rotational admittance, given by
\begin{equation}
\label{6.13}\mathcal{Y}_R(\omega)=[-i\omega I_p+\zeta_R(\omega)]^{-1}.
\end{equation}
We note that
\begin{equation}
\label{6.14}\lim_{\kappa\rightarrow\infty}\zeta_R(\omega)=8\pi\eta a^3\bigg[1+\alpha^2a^2\frac{6+\alpha a}{15(1+\alpha a)}\bigg].
\end{equation}
The last term in brackets differs from that for a sphere with no-slip boundary condition\cite{9}, where it is replaced by $\alpha^2a^2/(3+3\alpha a)$. In the latter case the added moment of inertia vanishes, but the friction coefficient grows linearly for large $\alpha a$, in contrast with the behavior shown in Eq. (6.14) for the limit $\kappa a\rightarrow\infty$. As in the case of translation, the behavior suggested by the impermeable limit Eq. (6.14) is misleading. The behavior of the no-slip sphere arises from just the first term in Eq. (6.2), denoted elsewhere\cite{11}$^,$\cite{11B} as $\vc{T}_\omega$.  In the present case the incoming angular momentum flow is used partly as a torque acting on the polymer, partly it is used to accelerate the fluid within the sphere of radius $a$.

We consider in particular the applied torque
\begin{equation}
\label{6.15}\vc{N}(t)=\vc{L}\;\delta(t),
\end{equation}
where $\vc{L}$ is the imparted angular momentum. Correspondingly $\vc{N}_\omega=\vc{L}$. We define the rotational velocity relaxation function $\gamma_R(t)$ by
\begin{equation}
\label{6.16}\Omega(t)=\frac{L}{I_p}\;\gamma_R(t)\qquad t>0.
\end{equation}
It has the properties
\begin{equation}
\label{6.17}\gamma_R(0+)=1,\qquad\int^\infty_0\gamma_R(t)\;dt=\frac{I_p}{\zeta_R(0)}.
\end{equation}

The relaxation function is related to the rotational admittance by
\begin{equation}
\label{6.18}\int^\infty_0e^{i\omega t}\gamma_R(t)\;dt=\mathcal{Y}_R(\omega).
\end{equation}
Hence the relaxation function $\gamma_R(t)$ can be evaluated by inverse Fourier transform.
At low frequency the admittance has the expansion
\begin{equation}
\label{6.19}\mathcal{Y}_R(\omega)=\mu_R(0)+i\big(I_p\mu_R(0)^2+y_{R2}\big)\omega+\frac{1}{24\pi\eta}\;\alpha^3+O(\omega^2),
\end{equation}
where $\mu_R(0)=\zeta_{R0}^{-1}$ with $\zeta_{R0}=\zeta_R(0)$ is the steady-state mobility, and $y_{R2}=-\zeta_{R2}/\zeta_{R0}^2$. The third term gives rise to the long-time behavior
\begin{equation}
\label{6.20}\vc{\Omega}(t)\approx\frac{1}{\pi^{3/2}\rho(4\nu t)^{5/2}}\;\vc{L}\qquad\mathrm{as}\;t\rightarrow\infty.
\end{equation}
This depends only on the properties of the fluid.

\section{\label{VII}Analysis of rotational velocity relaxation}

The analytic behavior of the admittance as a function of frequency can be analyzed in the same way as for translation. It is convenient to define the dimensionless admittance $\hat{F}_R(x)$ with the complex variable $x=\alpha a$ as
\begin{equation}
\label{7.1}\hat{F}_R(x)=\frac{8\pi I_p}{15I_f}\;\eta a^3\mathcal{Y}_R(\omega),
\end{equation}
where $I_f=8\pi\rho a^5/15$ is the moment of inertia of displaced fluid, and we have chosen the prefactor such that $\hat{F}_R(x)$ behaves as $1/x^2$ for large $x$. The function takes the form
\begin{equation}
\label{7.2}\hat{F}_R(x)=\frac{1}{x^2+MZ_R(x)},
\end{equation}
where we have abbreviated
\begin{equation}
\label{7.3}M=15I_f/I_p,\qquad\zeta_R(\omega)=8\pi\eta a^3 Z_R(x).
\end{equation}
From Eq. (6.7) we find that for small $x$ the resistance function $Z_R(x)$ has the expansion
\begin{equation}
\label{7.4}Z_R(x)=Z_0+Z_2 x^2-\frac{1}{3}Z_0^2 x^3+O(x^4),
\end{equation}
where $Z_0=Z_R(0)$ is given by Eq. (6.6) and $Z_2=\zeta_{R2}/(8 \pi\eta a^3)$ by Eq. (6.8).

We write the reduced admittance $\hat{F}_R(x)$ as a sum of simple poles in the complex $x$ plane, as in Eq. (4.4).
The decay as $1/x^2$ for large $x$ implies the sum rules Eq. (4.5). From the value at $x=0$ we find the sum rules Eq. (4.6).
In addition we find from Eq. (7.4)
\begin{equation}
\label{7.5}
\sum_j\frac{A_j}{x_j^3}=\frac{1+MZ_2}{M^2Z_0^2},\qquad\sum_j\frac{A_j}{x_j^4}=\frac{-1}{3M}.
\end{equation}
Finally, Eq. (6.10) implies the sum rules
\begin{equation}
\label{7.6}\sum_jA_jx_j^2=0,\qquad\sum_jA_jx_j^3=-MZ_{0\infty},\qquad Z_{0\infty}=\frac{1}{15}\kappa^2 a^2,
\end{equation}
where $Z_{0\infty}$ corresponds to $\zeta_{R\infty}$. The relaxation function $\gamma_R(t)$ is given by an expression of the form Eq. (4.8).

We write the function $\hat{F}_R(x)$ in the form
\begin{equation}
\label{7.7}\hat{F}_R(x)=\bigg[MZ_0+x^2+\frac{3MZ_2^2x^2}{3Z_2+Z_0^2x+x^2\psi(x)}\bigg]^{-1},
\end{equation}
with a function $\psi(x)$ which tends to a constant for $x\rightarrow 0$ and for $x\rightarrow\infty$. If we approximate $\psi(x)$ as a ratio of two polynomials of degree $d-4$, then $\hat{F}_R(x)$ is approximated by a Pad\'e approximant $\hat{F}_{RPd}(x)$ with $d$ poles in the complex $x$ plane. By construction the expansion of $\hat{F}_{RPd}(x)$ in powers of $x$ agrees with that of $\hat{F}_R(x)$ to terms of order $x^3$,
\begin{equation}
\label{7.8}\hat{F}_{RPd}(x)=\frac{1}{MZ_0}-\frac{1+MZ_2}{M^2Z_0^2}x^2+\frac{1}{3M}x^3+O(x^4),
\end{equation}
and the function behaves as $1/x^2$ as $x\rightarrow\infty$.

For large values of $\kappa a$ the steady-state friction coefficient tends to the Stokes value $8\pi\eta a^3$ for a no-slip solid sphere. It is therefore of interest to compare the relaxation function for large $\kappa a$ with that for a no-slip sphere. The latter follows from\cite{9}
\begin{equation}
\label{7.9}\hat{F}_{Rns}(x)=\bigg[M+x^2+\frac{M x^2}{3+3x}\bigg]^{-1},
\end{equation}
corresponding to $Z_0=1,\;Z_2=\frac{1}{3}$ and $\psi=0$. This function has three poles in the complex $x$ plane. For a sphere with mixed slip-stick boundary condition with a slip parameter $\xi$, taking the values $\xi=0$ for no-slip and $\xi=\frac{1}{3}$ for perfect slip, the function becomes\cite{21}
\begin{equation}
\label{7.10}\hat{F}_{Rss}(x)=\bigg[M(1-3\xi)+x^2+\frac{(1-3\xi)^2x^2}{3+3x+3\xi x^2}\bigg]^{-1},
\end{equation}
corresponding to $Z_0=1-3\xi,\;Z_2=\frac{1}{3}(1-3\xi)^2$ and $\psi=\xi(1-3\xi)^2$. In this case the function has four poles in the complex $x$ plane, except for $\xi=0$ and $\xi=\frac{1}{3}$.

The simplest Pad\'e approximant is obtained by neglecting the function $\psi(x)$ in Eq. (7.7). This yields
\begin{equation}
\label{7.11}\hat{F}_{RP3}(x)=\bigg[MZ_0+x^2+\frac{3MZ_2^2x^2}{3Z_2+Z_0^2x}\bigg]^{-1},
\end{equation}
a function with three poles in the complex $x$ plane. We call this the three-pole approximation. A higher order Pad\'e approximant can be found from values of the function $\hat{F}_R(x)$ at a small number of points on the positive $x$ axis.

In our numerical work we consider a neutrally buoyant polymer with $I_f=I_p$, so that $M=15$. We consider again three values of the parameter $\kappa a$, namely $\kappa a=0.2,\;\kappa a=1$, and $\kappa a=5$. For these three values we find for the poles of $\hat{F}_{RP3}(x)$
\begin{eqnarray}
\label{7.12}x_0&=&-1.428,\qquad x_\pm=-7\times 10^{-7}\pm0.200i,\qquad\kappa a=0.2,\nonumber\\
x_0&=&-1.450,\qquad x_\pm=-0.006\pm0.947i,\qquad\kappa a=1,\nonumber\\
x_0&=&-2.436,\qquad x_\pm=-0.708\pm1.962i,\qquad\kappa a=5,
\end{eqnarray}
showing considerable variation as a function of $\kappa a$. The corresponding amplitudes are
\begin{eqnarray}
\label{7.13}A_0&=&-3\times10^{-5},\qquad A_\pm=-10^{-5}\mp2.505i,\qquad\kappa a=0.2,\nonumber\\
A_0&=&-0.009,\qquad A_\pm=0.004\mp0.521i,\qquad\kappa a=1,\nonumber\\
A_0&=&-0.158,\qquad A_\pm=0.079\mp0.185i,\qquad\kappa a=5.
\end{eqnarray}
For comparison we have for the no-slip sphere
\begin{eqnarray}
\label{7.14}x_0&=&-2.322,\qquad x_\pm=-1.839\pm1.754i,\nonumber\\
A_0&=&-0.399,\qquad A_\pm=0.200\mp0.230i,\qquad\mathrm{no-slip}.
\end{eqnarray}
Comparing the function $\hat{F}_{RP3}(x)$ with the exact $\hat{F}_R(x)$ for positive values of $x$ we find that for $\kappa a=0.2$ the functions are nearly identical, that for $\kappa a=1$ they differ at most by a few promille, but that for $\kappa a=5$ the ratio differs from unity by more than seven percent. This indicates that for values up to $\kappa a=1$ the three-pole approximation is excellent, but for $\kappa a>1$ more poles must be taken into account.

In the next approximation $\psi(x)$ in Eq. (7.7) is set equal to a constant $\psi_0$. This yields a four-pole approximation $\hat{F}_{RP4}(x)$. The mixed-slip expression Eq. (7.10) is of this form with $\psi_0=\xi(1-3\xi)^2$, as noted above. In the problem of viscoelasticity of a colloidal suspension\cite{22}, where the mathematical structure is the same, we have recommended to determine the constant $\psi_0$ from the exact value of $\hat{F}_R(x)$ at $x=\sqrt{\tau_v/\tau_M}=\sqrt{MZ_0}$. This corresponds to $\psi_0=0.0009$ for $\kappa a=1$, and to $\psi_0=0.0034$ for $\kappa a=5$. If we compare $\hat{F}_{RP4}(x)$, thus determined, with the exact $\hat{F}_R(x)$, we see that for $\kappa a=1$ the agreement is improved by an order of magnitude, but for for $\kappa a=5$ there is hardly any change. This indicates that the required number of poles grows with $\kappa a$.

In Fig. 6 we show $\log_{10}\gamma_R(t)$ as a function of $\log_{10}(t/\tau_v)$ for the three different values of $\kappa a$, and compare with the no-slip function $\gamma_{Rns}(t)$. The function $\gamma_R(t)$ decays more slowly than the no-slip function, corresponding to the smaller value of the steady-state friction coefficient. At $\kappa a=5$ the ratio is $Z_0(5)=0.520$. At $t=2\tau_v$ the functions have decayed to $\gamma_R(2\tau_v)=0.0021$ and $\gamma_{Rns}(2\tau_v)=0.0013$. In the long-time regime the two functions become identical.

The spectral density $p_R(u)$ is defined in the same way as in the translational case, and can be calculated as in Eq. (4.20).
For small values of $\kappa a$ it is given by a sharp peak. Then the function $\hat{F}_R(x)$ is well approximated by a two-pole expression with two conjugate poles $x_\pm$ with small negative real part, as exemplified for $\kappa a=0.2$ in Eqs. (7.12) and (7.13). In this case the relaxation function is nearly exponential with relaxation time $\tau_M=\tau_v/(MZ_0)$. At $\kappa a=0.2$ this corresponds to $\sqrt{\tau_v/\tau_M}=0.200$. For larger values of $\kappa a$ the spectral density broadens, and the decay function is more complicated. In Fig. 7 we show the spectral density for $\kappa a=1$ and compare with the four-pole approximation corresponding to $\psi_0=0.0009$. The peak has broadened, and the four-pole approximation is accurate. In Fig. 8 we show the spectral density for $\kappa a=5$ and compare with the four-pole approximation corresponding to $\psi_0=0.0034$. In this case there is an additional sharp peak in the spectrum, which is not present in the four-pole approximation. In Fig. 9 we show the ratio of the two relaxation functions as a function of $t/\tau_v$. At $t=4\tau_v$ the exact function has decayed to $324\times 10^{-6}$ and the four-pole function to $337\times 10^{-6}$. At long times the two functions become identical.

The short-time behavior corresponding to the peak in the relaxation spectrum for large $\kappa a$ is complicated. In Fig. 10 we show the ratio $\gamma_R(t)/\gamma_{Rns}(t)$ for $\kappa a=10$ in a short time-interval. The ratio is at first larger than unity, then less, before rising to $1.850$ at $t=0.5\tau_v$, and finally slowly decaying to unity. We note that $Z_0(10)=0.730$, so that the overall timescale, as given by the integral of the relaxation function, for the permeable sphere is a factor 1.37 larger than that of the no-slip sphere. The short-time behavior of the flow velocity is similar to that of the no-slip sphere. In both cases the flow velocity immediately after the twist vanishes, $\vc{v}(\vc{r},0+)=0$. In the case of the permeable sphere the flow then builds up due to friction caused by relative motion between fluid and rigid polymer. In the case of the no-slip sphere the flow is caused by stress exerted at the sphere surface. The explicit expression for the Fourier transform of the flow velocity about the no-slip sphere, given by Felderhof and Jones\cite{11}, agrees with that given by Eqs. (5.8) and (5.9) in the limit of large $\kappa a$. Hence one sees that for any $r>a$ the flow velocity vanishes initially, and then builds up before decaying to zero.

\section{\label{V}Discussion}

We have shown that the translational and rotational velocity relaxation functions of a permeable sphere have interesting behavior. In particular, the vanishing of the added mass is remarkable. As a consequence, in the short and intermediate time regime the translational velocity relaxation function differs strongly from that of a no-slip sphere. The relaxation spectrum has corresponding remarkable features. Similarly, the rotational velocity relaxation function of a permeable sphere differs strongly from that of a no-slip sphere. At long times both the translational and rotational velocity relaxation function show the well-known universal behavior with an algebraic long-time tail, identical to that of a no-slip sphere.

Via the fluctuation-dissipation theorem\cite{23} the velocity relaxation functions are directly related to the velocity autocorrelation functions of Brownian motion\cite{18}$^,$\cite{23A}. Presently the translational autocorrelation function can be studied experimentally in detail\cite{24}. It would be of interest to apply the same experimental methods to the Brownian motion of a spherically symmetric polymer. Alternatively the Brownian motion of a permeable sphere could be studied in computer simulation, and be compared with that of a no-slip solid sphere.

In the above we have considered only the simplest polymer model. Clearly the same theoretical method can be applied to a spherical shell model\cite{25}, a coated sphere model\cite{26}$^,$\cite{27}, or some other radially symmetric permeability profile. In cases where the differential equations cannot be solved analytically, for example for a gaussian profile, the Pad\'e approximant method of Sec. 4 would be a necessary ingredient. For translational motion it would suffice to solve the differential equations Eq. (2.8) numerically for a small number of positive values $\alpha a$, and fit to the outside flow of the form Eq. (2.13). Similarly, for rotational motion it would suffice to solve the differential equation Eq. (5.5) for a small number of positive values $\alpha a$, and fit to the outside flow of the form Eq. (5.8).

\newpage
\appendix
\section{}
The vector spherical harmonics $\vc{A}_{lm}, \vc{B}_{lm}$ and $\vc{C}_{lm}$ are defined by\cite{9}
\begin{eqnarray}
\label{A.1}\vc{A}_{lm}(\theta,\varphi)&=&lY_{lm}\vc{e}_r+\frac{\partial Y_{lm}}{\partial\theta}\;\vc{e}_\theta
+\frac{1}{\sin\theta}\frac{\partial Y_{lm}}{\partial\varphi}\;\vc{e}_\varphi,\nonumber\\
\vc{B}_{lm}(\theta,\varphi)&=&-(l+1)Y_{lm}\vc{e}_r+\frac{\partial Y_{lm}}{\partial\theta}\;\vc{e}_\theta
+\frac{1}{\sin\theta}\frac{\partial Y_{lm}}{\partial\varphi}\;\vc{e}_\varphi,\nonumber\\
\vc{C}_{lm}(\theta,\varphi)&=&\frac{1}{\sin\theta}\frac{\partial Y_{lm}}{\partial\varphi}\;\vc{e}_\theta
-\frac{\partial Y_{lm}}{\partial\theta}\;\vc{e}_\varphi,
\end{eqnarray}
where $Y_{lm}(\theta,\varphi)$ are spherical harmonics in the notation of Edmonds\cite{28} and $\vc{e}_r,\;\vc{e}_\theta,\;\vc{e}_\varphi$ are unit vectors
in spherical coordinates. Apart from normalization the vector
spherical harmonics are identical to the $\vc{Y}_{JlM}$ defined by
Edmonds. In this Appendix we list some identities which are useful in the analysis of Eqs. (2.3) and (2.5). We note that $\vc{e}_z$ is proportional to $\vc{A}_{10}$, and $(\vc{1}-3\hat{\vc{r}}\hat{\vc{r}})\cdot\vc{e}_z$ is proportional to $\vc{B}_{10}$.

The gradient formula reads
\begin{equation}
\label{A.2}\nabla(f(r)Y_{lm})=\frac{1}{2l+1}\bigg[\big[f'(r)+(l+1)\frac{f(r)}{r}\big]\vc{A}_{lm}+\big[-f'(r)+l\frac{f(r)}{r}\big]\vc{B}_{lm}\bigg].
\end{equation}
The curl operation yields
\begin{eqnarray}
\label{A.3}\nabla\times(f(r)\vc{A}_{lm})&=&\big[-f'(r)+(l-1)\frac{f(r)}{r}\big]\vc{C}_{lm},\nonumber\\
\nabla\times(f(r)\vc{B}_{lm})&=&\big[-f'(r)-(l+2)\frac{f(r)}{r}\big]\vc{C}_{lm},\nonumber\\
\nabla\times(f(r)\vc{C}_{lm})&=&\frac{l+1}{2l+1}\big[f'(r)+(l+1)\frac{f(r)}{r}\big]\vc{A}_{lm}+\frac{l}{2l+1}\big[f'(r)-l\frac{f(r)}{r}\big]\vc{B}_{lm}.
\end{eqnarray}
The divergence operation yields
\begin{eqnarray}
\label{A.4}\nabla\cdot(f(r)\vc{A}_{lm})&=&l\big[f'(r)-(l-1)\frac{f(r)}{r}\big]Y_{lm},\nonumber\\
\nabla\cdot(f(r)\vc{B}_{lm})&=&-(l+1)\big[f'(r)+(l+2)\frac{f(r)}{r}\big]Y_{lm},\nonumber\\
\nabla\cdot(f(r)\vc{C}_{lm})&=&0.
\end{eqnarray}
The Laplace operator yields
\begin{eqnarray}
\label{A.5}\nabla^2(f(r)\vc{A}_{lm})&=&\big[f''(r)+\frac{2}{r}f'(r)-l(l-1)\frac{f(r)}{r^2}\big]\vc{A}_{lm},\nonumber\\
\nabla^2(f(r)\vc{B}_{lm})&=&\big[f''(r)+\frac{2}{r}f'(r)-(l+1)(l+2)\frac{f(r)}{r^2}\big]\vc{B}_{lm},\nonumber\\
\nabla^2(f(r)\vc{C}_{lm})&=&\big[f''(r)+\frac{2}{r}f'(r)-l(l+1)\frac{f(r)}{r^2}\big]\vc{C}_{lm}.
\end{eqnarray}

\newpage

\section*{Figure captions}

\subsection*{Fig. 1}
Plot of the logarithm of the relaxation function $\log_{10}\gamma_T(t)$ as a function of $\log_{10}(t/\tau_v)$ for a neutrally buoyant sphere with $\kappa a=0.2$ (short dashes), $\kappa a= 1$ (long dashes), and $\kappa a=5$ (solid curve). We compare with  $\log_{10}\gamma_{Tns}(t)$ for a neutrally buoyant no-slip sphere of the same radius (dash-dotted curve).

\subsection*{Fig. 2}
Plot of the ratio of the relaxation function $\gamma_T(t)$ for $\kappa a=5$ to the no-slip function $\gamma_{Tns}(t)$ as a function of $t/\tau_v$.

\subsection*{Fig. 3}
Plot of the spectral density $p_T(u)$ as a function of $\log_{10}u$ for $\kappa a=5$ (solid curve), compared with the spectral density $p_{Tns}(u)$ for the no-slip sphere, as given by Eq. (4.21) (dashed curve). We also plot the spectral $p_{TP6}(u)$ obtained from a 6-pole Pad\'e approximation. The latter curve can be distinguished from the solid one only at the sharp peak.

\subsection*{Fig. 4}
Plot of the ratio of relaxation functions $\gamma_T(t)$ and $\gamma_{TP6}(t)$ as a function of $\log_{10}(t/\tau_v)$ for a permeable sphere with $\kappa a=5$. This shows that the six-pole Pad\'e approximant is quite accurate.

\subsection*{Fig. 5}
Plot of the spectral density $p_T(u)$ for $\kappa a=0.2$ as a function of $\log_{10}u$ (solid curve), compared with the approximate function $p_{TP2}(u)$ found in two-pole approximation, as given by Eq. (4.22) (dashed curve).

\subsection*{Fig. 6}
Plot of the logarithm of the relaxation function $\log_{10}\gamma_R(t)$ as a function of $\log_{10}(t/\tau_v)$ for a neutrally buoyant sphere with $\kappa a=0.2$ (short dashes), $\kappa a= 1$ (long dashes), and $\kappa a=5$ (solid curve). We compare with  $\log_{10}\gamma_{Rns}(t)$ for a neutrally buoyant no-slip sphere of the same radius (dash-dotted curve).

\subsection*{Fig. 7}
Plot of the spectral density $p_R(u)$ for $\kappa a=1$ as a function of $\log_{10}u$ (solid curve), compared with the approximate function $p_{TP4}(u)$ found in four-pole approximation with $\psi_0=0.0009$ (dashed curve). The two curves cannot be distinguished on the scale of the figure.

\subsection*{Fig. 8}
Plot of the spectral density $p_R(u)$ for $\kappa a=5$ as a function of $\log_{10}u$ (solid curve), compared with the approximate function $p_{TP4}(u)$ found in four-pole approximation with $\psi_0=0.0034$ (dashed curve).

\subsection*{Fig. 9}
Plot of the ratio of relaxation functions $\gamma_R(t)$ and $\gamma_{RP4}(t)$ as a function of $t/\tau_v$ for a permeable sphere with $\kappa a=5$.

\subsection*{Fig. 10}
Plot of the ratio of relaxation functions $\gamma_R(t)$ and $\gamma_{Rns}(t)$ as a function of $t/\tau_v$ for a permeable sphere with $\kappa a=10$.

\newpage
\setlength{\unitlength}{1cm}
\begin{figure}
 \includegraphics{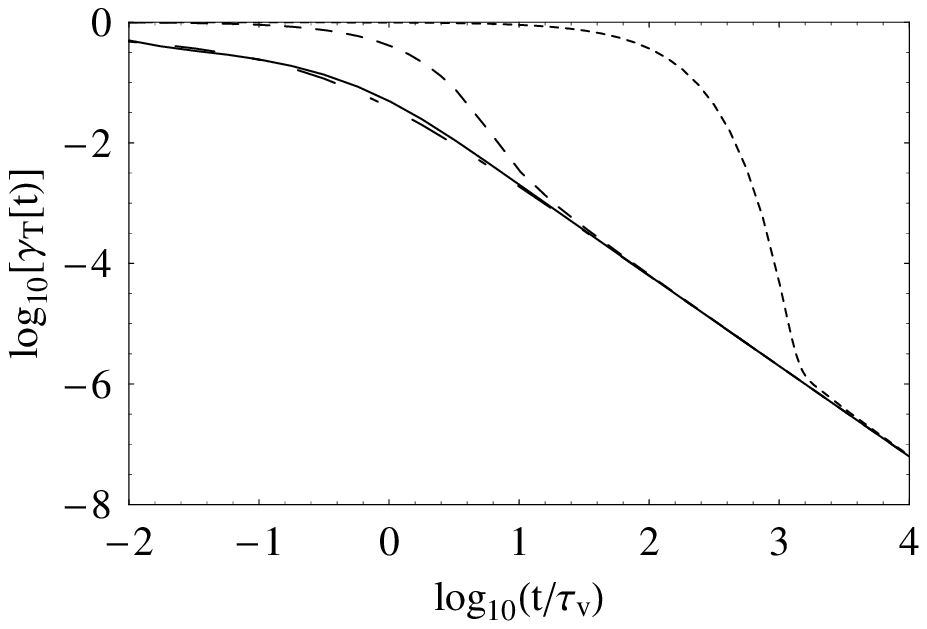}
   \put(-9.1,3.1){}
\put(-1.2,-.2){}
  \caption{}
\end{figure}
\newpage
\clearpage
\newpage
\setlength{\unitlength}{1cm}
\begin{figure}
 \includegraphics{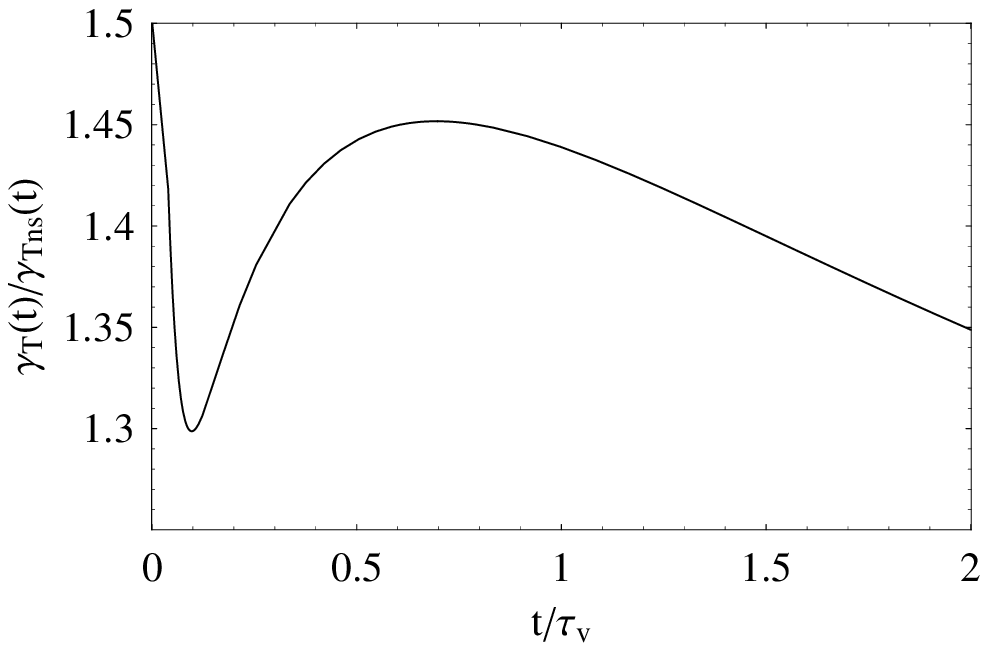}
   \put(-9.1,3.1){}
\put(-1.2,-.2){}
  \caption{}
\end{figure}
\newpage
\clearpage
\newpage
\setlength{\unitlength}{1cm}
\begin{figure}
 \includegraphics{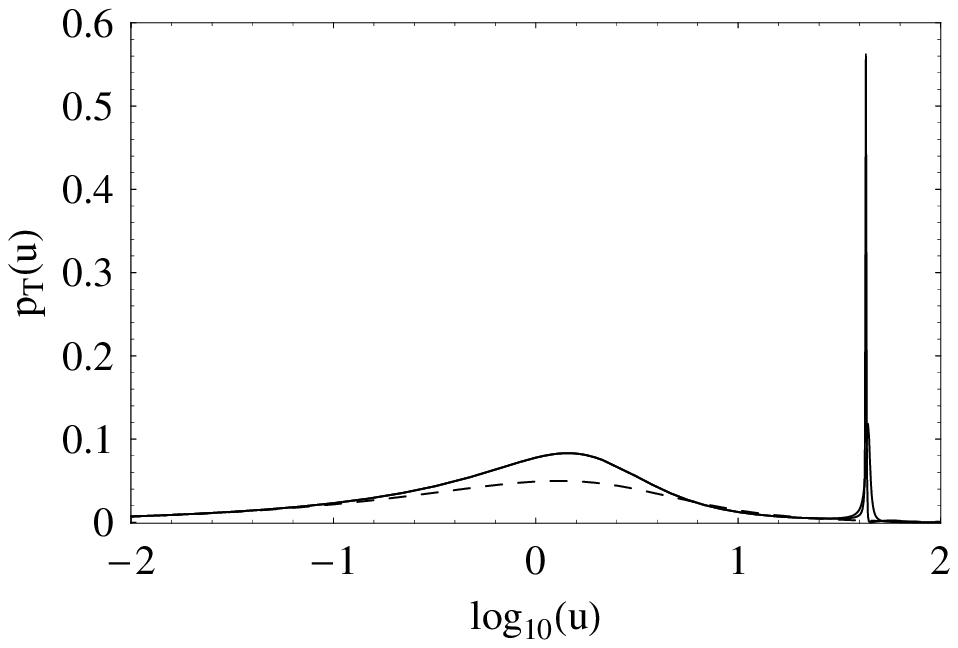}
   \put(-9.1,3.1){}
\put(-1.2,-.2){}
  \caption{}
\end{figure}
\newpage
\clearpage
\newpage
\setlength{\unitlength}{1cm}
\begin{figure}
 \includegraphics{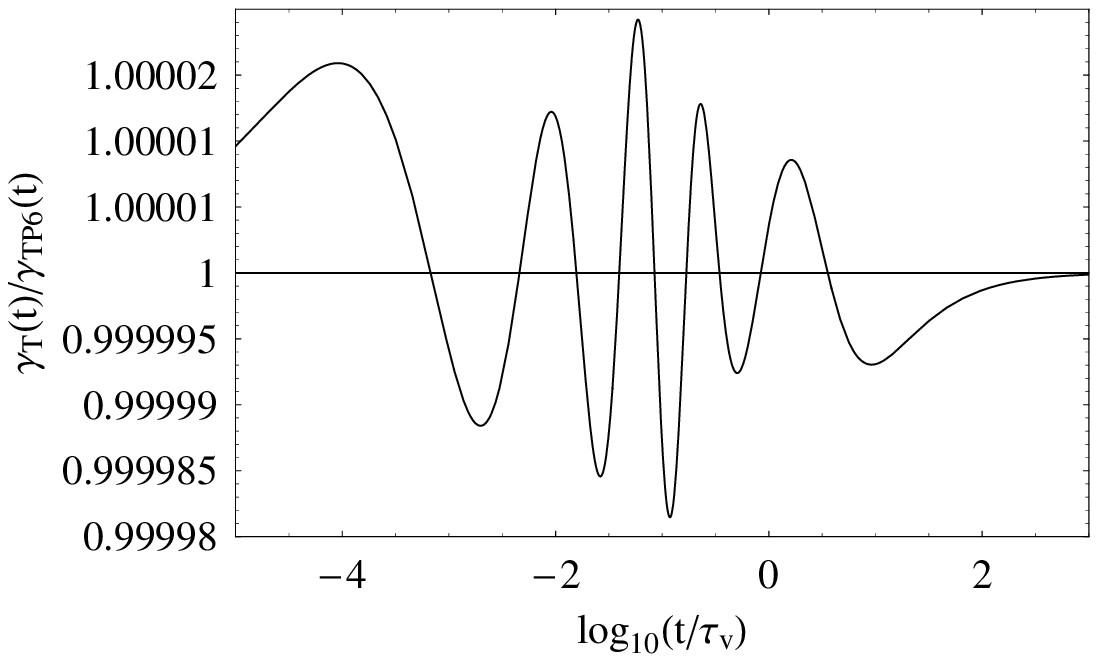}
   \put(-9.1,3.1){}
\put(-1.2,-.2){}
  \caption{}
\end{figure}
\newpage
\clearpage
\newpage
\setlength{\unitlength}{1cm}
\begin{figure}
 \includegraphics{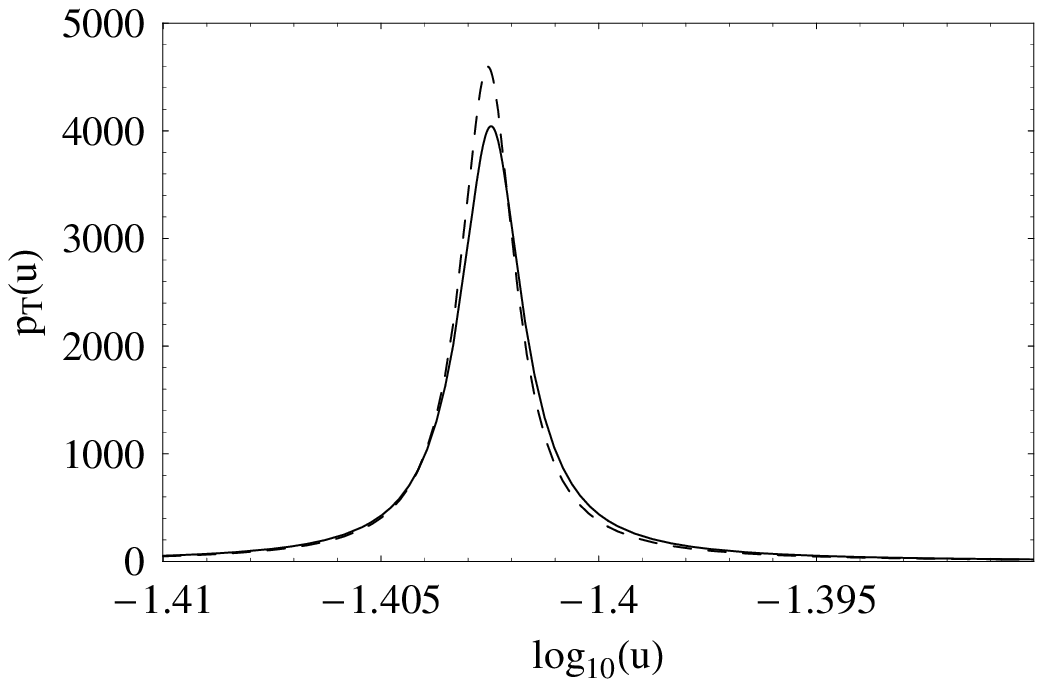}
   \put(-9.1,3.1){}
\put(-1.2,-.2){}
  \caption{}
\end{figure}
\newpage
\clearpage
\newpage
\setlength{\unitlength}{1cm}
\begin{figure}
 \includegraphics{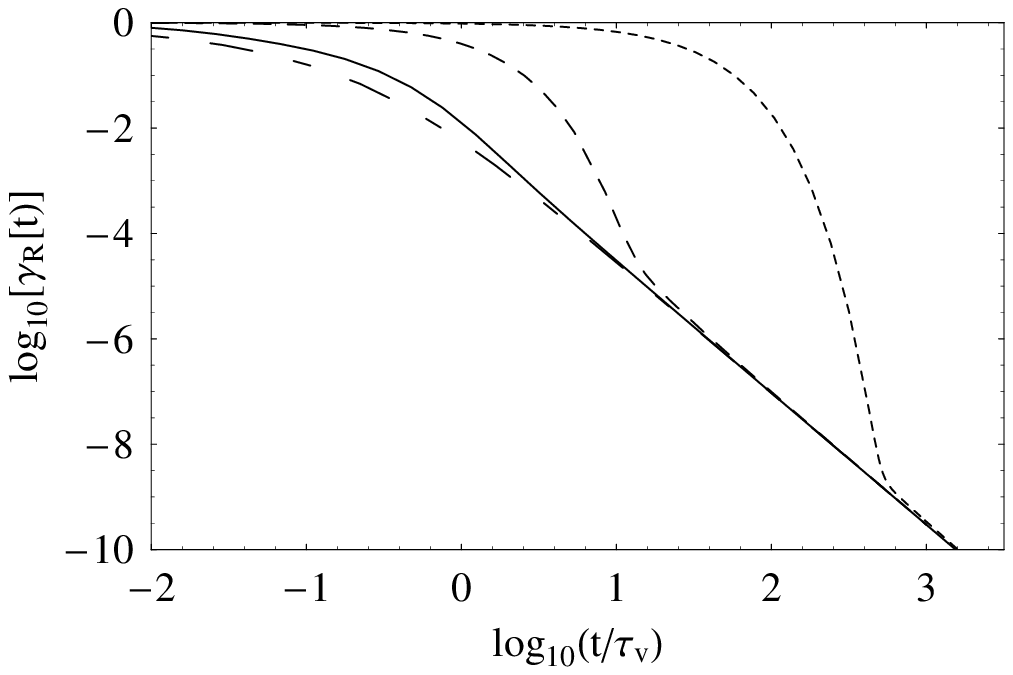}
   \put(-9.1,3.1){}
\put(-1.2,-.2){}
  \caption{}
\end{figure}
\newpage
\clearpage
\newpage
\setlength{\unitlength}{1cm}
\begin{figure}
 \includegraphics{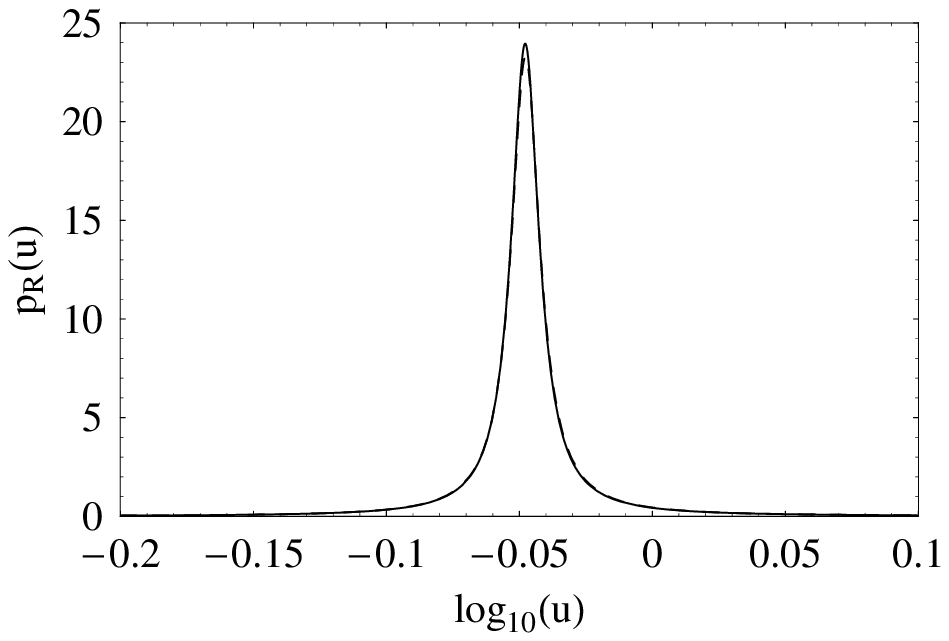}
   \put(-9.1,3.1){}
\put(-1.2,-.2){}
  \caption{}
\end{figure}
\newpage
\clearpage
\newpage
\setlength{\unitlength}{1cm}
\begin{figure}
 \includegraphics{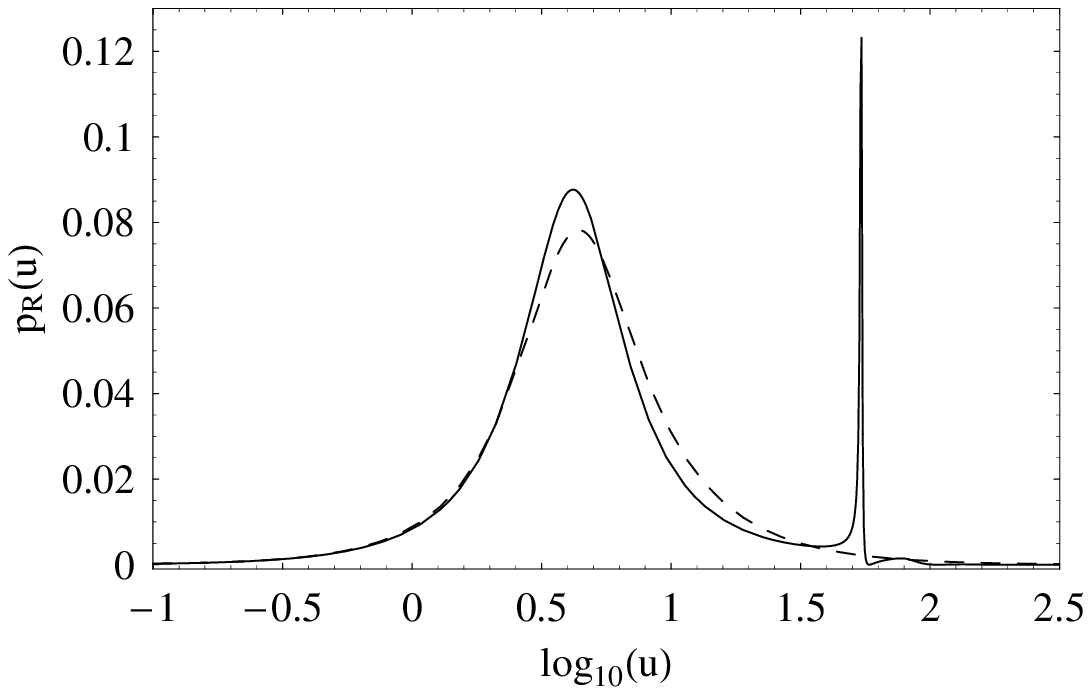}
   \put(-9.1,3.1){}
\put(-1.2,-.2){}
  \caption{}
\end{figure}
\newpage
\clearpage
\newpage
\setlength{\unitlength}{1cm}
\begin{figure}
 \includegraphics{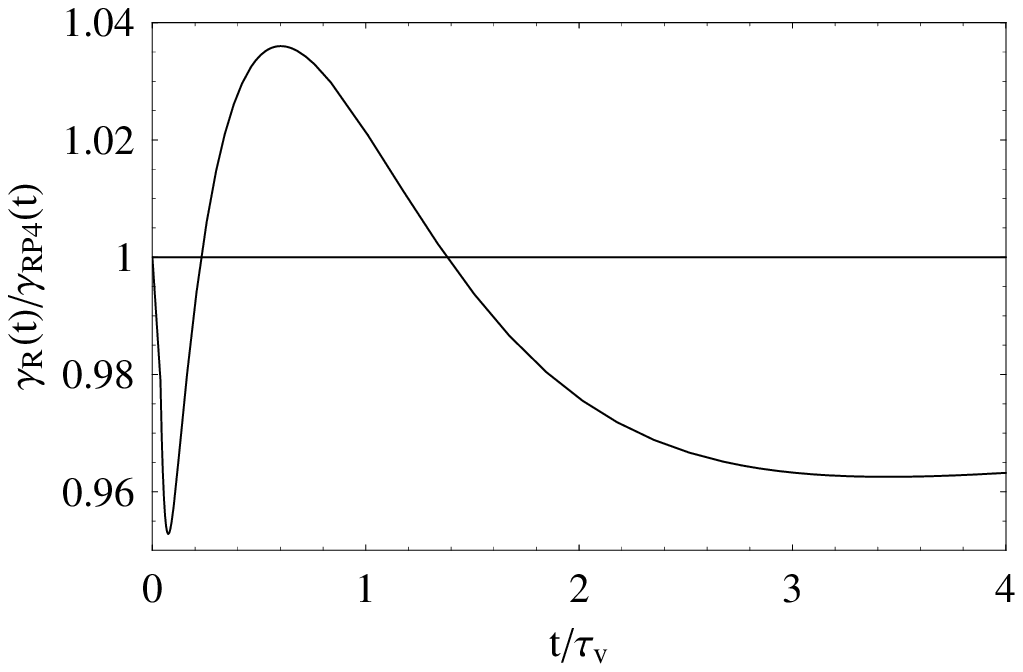}
   \put(-9.1,3.1){}
\put(-1.2,-.2){}
  \caption{}
\end{figure}
\newpage
\clearpage
\newpage
\setlength{\unitlength}{1cm}
\begin{figure}
 \includegraphics{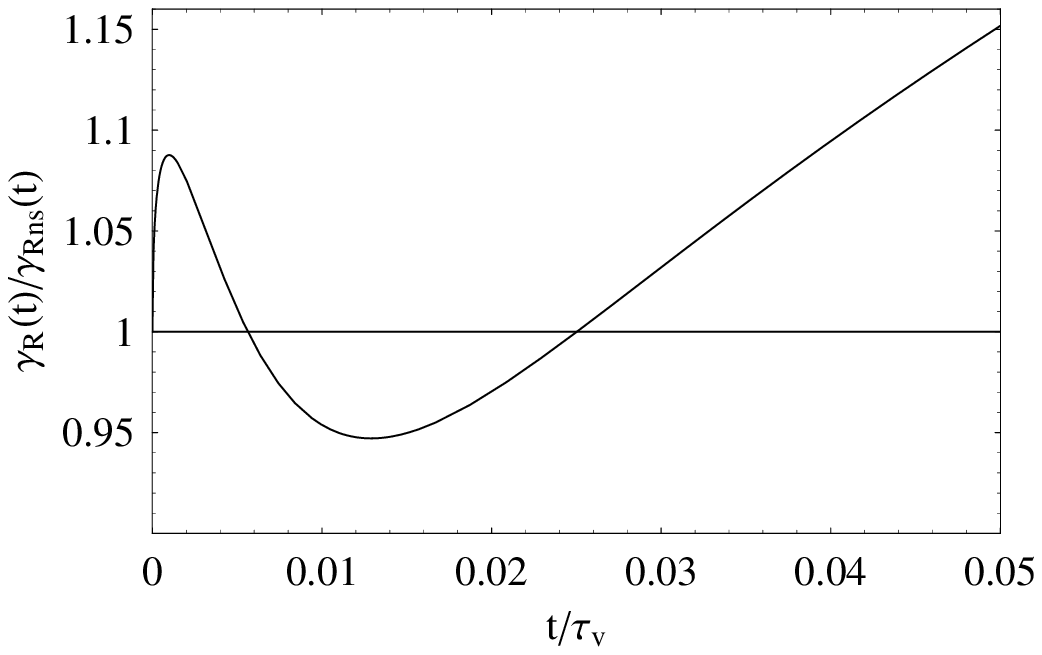}
   \put(-9.1,3.1){}
\put(-1.2,-.2){}
  \caption{}
\end{figure}
\newpage
\end{document}